\documentclass[11pt,twoside]{article}

\usepackage{asp2006}
\usepackage{graphicx}
\usepackage{epsf}

\begin {document}

\title{Boxy/Peanut bulges and stellar bars}

\begin{quote}
Inma Martinez-Valpuesta$^{1,2}$ and E. Athanassoula$^2$\\
{\itshape $^1$Instituto de Astrof\'\i sica de Canarias, E-38200 La Laguna, Tenerife, Spain}\\
{\itshape $^2$LAM, Observatoire Astronomique de Marseille Provence, 2 Place Le Verrier, F-13004 Marseille, France}\\
\end{quote}


\begin{abstract}
Boxy/peanut bulges in disc galaxies have been associated to stellar
bars. We analyse their properties in a large sample of
$N$-body simulations, using different methods to measure their 
strength, shape and possible asymmetry, and then
inter-compare the results. Some of these methods can be applied to both
simulations and observations. In particular, we seek correlations
between bar and peanut properties, which, when applied to real
galaxies, will give information on bars in edge-on galaxies, and
on peanuts in face-on galaxies.
\end{abstract}

\section{Introduction}
Both simulations and theoretical studies have 
shown that bars are not vertically thin morphological
features, but have a considerable vertical extent and a 
vertical structure, known as the Boxy/Peanut bulges (hereafter B/P;
Combes \& Sanders 1981, Combes et al. 1990). Comparisons between
observations and $N$-body simulations have established this
direct connection firmer (Athanassoula 2005 and references
therein). Observations have shown that both bars and B/P bulges are
quite predominant in disc galaxies and that the corresponding
frequencies are in good agreement with the link between the two
structures (L\"utticke, Dettmar \& Pohlen 2000). 

We measure the peanut properties in a large sample of several hundred
$N$-body simulations ran by one of us (EA) for different
purposes. More information on these simulations and on their
properties can be found in Athanassoula \& Misiriotis (2002) and
Athanassoula (2003, 2007). In particular, we seek correlations between
the properties of the bar and the properties of the B/P bulge. 

\section{Methods for measuring bar and peanut strength and correlations}
In order to measure the bar strength, we use standard Fourier
decomposition and take the amplitude of the second ($m$ = 2)
component. 

\begin{equation}
C_{m,r} (R) = |\sum_{j=1}^{N_s}~m_j e^{i m \theta_j}|, ~~~~m=2,
\label{eq:Cmr}
\end{equation}

\noindent
where $m_j$ and $\theta_j$ are the mass and azimuthal angle of
particle $j$. This can be done either globally or as a function of radius. In the
former case the summation is carried out over all particles in the
disc, while in the latter $N_s$ is the number of particles in a
given cylindrical shell of radius $R$ and $C_{m,r} = C_{m,r} (R)$. 

The vertical asymmetry and the strength of the B/P can be measured in
a similar way, by  

\begin{equation}
C_{m,z} = | \sum_{j=1}^{N_s}~z_j e^{i m z_j / (5 z_0)}|, ~~~~~~~m=1, 2,
\label{eq:Cmz}
\end{equation}

\noindent
where $N_s$ can be either the number of particles in the disc
component, or the number of particles per vertical column or cut (in
which case $C_{m,z} = C_{m,z} (R)$) and $z_0$ is the scale height of the
initial exponential disc. In Fig.~1 we apply these two definitions to
two simulation snapshots, one with a boxy bulge and the other with a
peanut, or X-shaped bulge.  

\begin{figure}
\vspace{0.1in}
\begin{center}
\includegraphics[angle=0.0,scale=0.75]{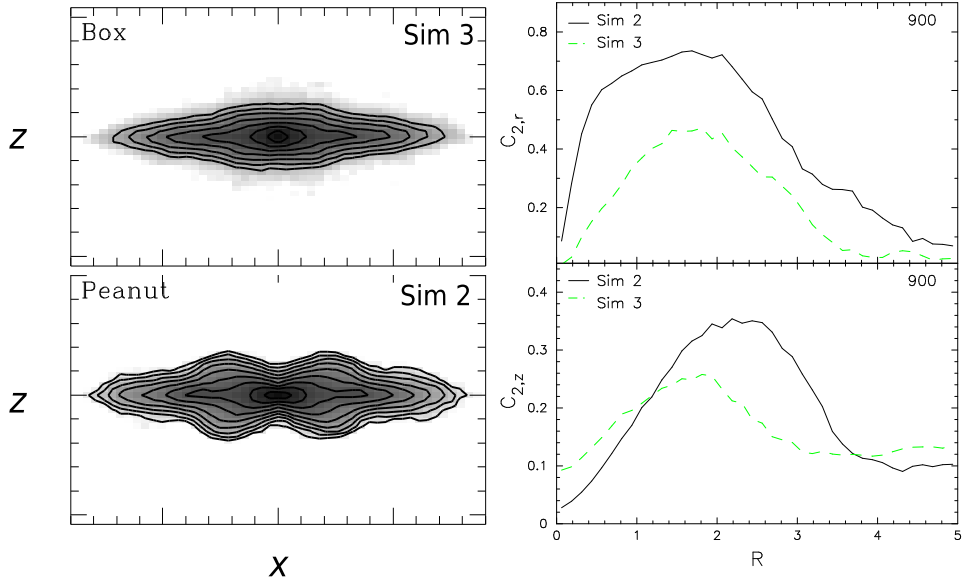}
\includegraphics[trim= 0 0 -120 0,origin=52 51 479 399,angle=-90,scale=0.5]{Martinez-Valpuesta_fig2.eps}
\vspace{-0.8in}
\caption{{\it Right figure}: Strength of the bar (top panel) and
  of the B/P bulge (lower panel) vs. radius, for both snapshots. {\it
  Left figure}: Edge-on, side-on view of the boxy (top panel) and of
  the strong peanut (lower panel) bulges used in the right panels.
}
\end{center}
\vspace{-0.3in}
\end{figure}

We have also applied statistical methods, based on the distribution of
the $z$ coordinates of the particles in a given vertical cut
(perpendicular to the equatorial 
plane) of a snapshot seen edge-on, with the bar viewed side-on. We use
the median of the absolute values and the standard deviation ($\sigma_z$) 
to measure the strength of the peanut, the kurtosis to measure its shape
and the skewness and mean to measure the strength of the buckling
event i.e. the asymmetry during the vertical instability.   
 
Our methods to calculate the strength of the B/P bulge correlate
nicely (Fig.~2, {\it right panels}) and can also be applied to real,
highly inclined galaxies. Even more interesting, we find correlations
between the strength and shape of the B/P bulge with the strength of the
bar (Fig.~3). Note in this figure how some simulations undergo several
buckling events, as previously noted by Martinez-Valpuesta, Shlosman
\& Heller (2006). 

\begin{figure}
\vspace{0.1in}
\begin{center}
\includegraphics[scale=0.8]{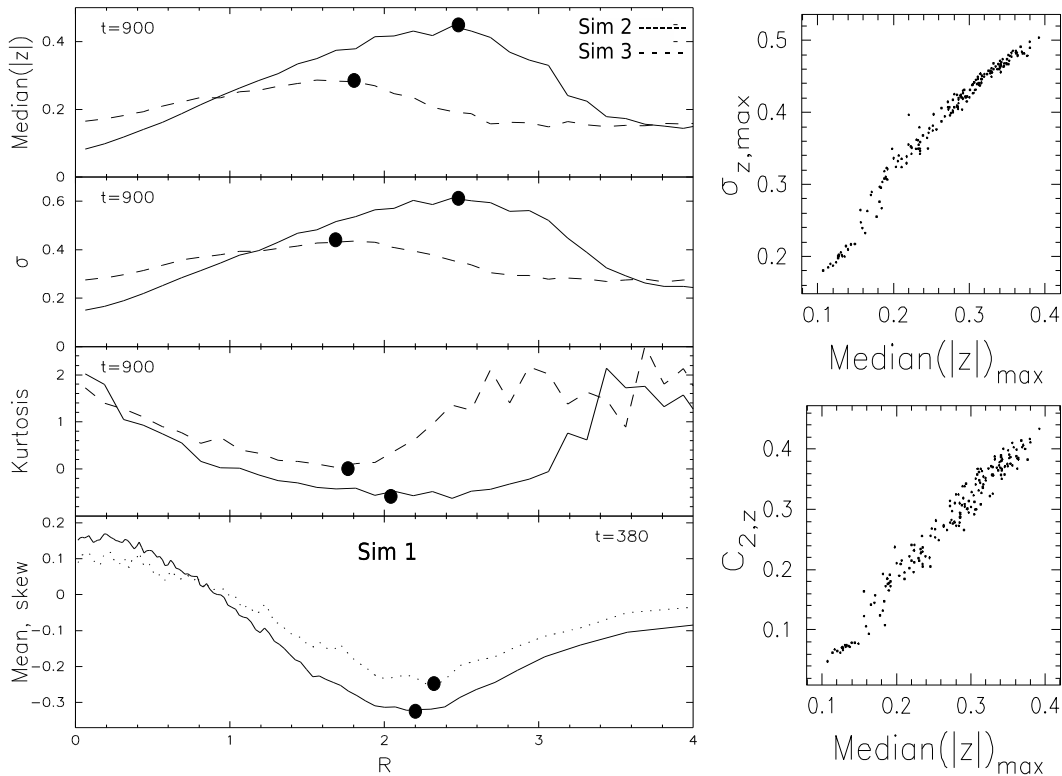}
\caption{{\it Left panels}: Results of the statistical methods vs. radius. The
  upper three panels are for the boxy and peanut snapshots shown in
  Fig. 1 and the lower one for a snapshot caught at the time of the
  buckling. {\it Right panels}: Two correlations showing how the
  statistical moments method agrees with the Fourier analysis method. 
}
\end{center}
\vspace{-0.1in}
\end{figure}

\section{Conclusions}

We presented several methods to calculate the strength of the bar and
the strength, shape and asymmetry of the B/P bulge and found strong
correlations between their results. The most important correlation
relates the strength of the bar with the strength of the B/P bulge,
the strongest bars having the strongest peanuts. We also find that the
strength of the peanut depends on the number of buckling episodes it
underwent, the strongest bars having undergone more buckling episodes
(Fig. 3). Finally, we find a very interesting result about $C_{2,z}(R)$,
i.e. about the shape of the radial density profiles along cuts
perpendicular to the equatorial plane. For strong bars, having a
strong peanut or X-shaped bulge, this profile is more flat-topped,
while for weaker bars, with more boxy-like bulges, it is more
peaked. All the results summarized here are
discussed in length by Athanassoula \& Martinez-Valpuesta (2008, in
preparation). 

\begin{figure}
\begin{center}
\includegraphics[scale=0.8]{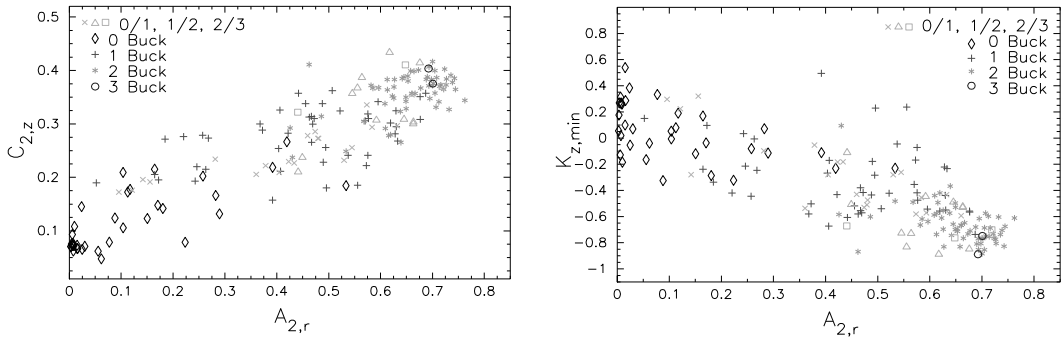}
\caption{Correlations between bar and peanut properties. 
Each symbol corresponds to one simulation. The type of symbol is
related to the number of buckling events suffered by the bar during
its evolution. {\it Left panel}: Strength of the B/P bulge measured
with our Fourier based method vs. the strength of the bar. {\it Right
  panel}: Shape of the B/P bulge (i.e. shape of the radial density
profiles along cuts perpendicular to the equatorial plane, measured by
the minimum of the kurtosis) plotted as a function of bar strength. 
}
\end{center}
\end{figure}
\acknowledgements This work has been partially supported by the Peter Gruber Foundation Fellowship and by grant ANR-06-BLAN-0172

\end{document}